# Pressure-tuned double-dome superconductivity in KZnBi with honeycomb lattice

Cuiying Pei[1#], Hongjoo Ha[2#], Sen Shao[3,4#], Shihao Zhu[1], Qi Wang[1,5], Juefei Wu[1], Yanchao Wang[3,4], Yulin Chen[1,5,6], Yanming Ma[3,4,7]*, Sung Wng Kim[2]*, Yanpeng Qi[1,5,8]*

1. State Key Laboratory of Quantum Functional Materials, School of Physical Science and Technology, ShanghaiTech University, Shanghai 201210, China
2. Department of Energy Science, Sungkyunkwan University, Suwon 16419, Republic of Korea
3. Key Laboratory of Material Simulation Methods & Software of Ministry of Education, College of Physics, Jilin University, Changchun 130012, China
4. State Key Laboratory of Superhard Materials, College of Physics, Jilin University, Changchun 130012, China
5. ShanghaiTech Laboratory for Topological Physics, ShanghaiTech University, Shanghai 201210, China
6. Department of Physics, Clarendon Laboratory, University of Oxford, Parks Road, Oxford OX1 3PU, UK
7. International Center of Future Science, Jilin University, Changchun 130012, China
8. Shanghai Key Laboratory of High-resolution Electron Microscopy, ShanghaiTech University, Shanghai 201210, China

\# These authors contributed to this work equally.

\* Correspondence should be addressed to Y.Q. (qiyp@shanghaitech.edu.cn), S.K. (kimsungwng@skku.edu) and Y.M. (mym@jlu.edu.cn).

Keywords: superconductivity, high pressure, honeycomb lattice

**Abstract**

Materials with honeycomb lattice structures exhibit unique electronic properties arising from their distinctive atomic arrangements. Their weakly coupled nature facilitates modulation by external stimuli, which leads to a diverse range of physical phenomena, particularly superconductivity. Here, we report the discovery of a pressure-induced M-shaped double-dome superconducting phase in KZnBi with honeycomb lattice. Under applied pressure, the superconducting transition temperature $T_c$ increases sharply and reaches a maximum value of 7 K at approximately 2.5 GPa. Following a structural phase transition from the ambient-pressure $P6_3/mmc$ phase to the high-pressure $Pnma$ phase, $T_c$ gradually decreases. Further compression induces an electronic transition near 7 GPa, accompanied by an unexpected reentrant superconducting phase with a higher $T_c$ of 8 K. Our theoretical calculations indicate that KZnBi undergoes a transition from a Dirac band structure to a strong topological semimetal state following the structural phase transition. These findings establish KZnBi as an ideal platform for investigating the diverse structural manifestations and intrinsic phenomena of the honeycomb lattice, demonstrating the fundamental importance of honeycomb structures in advancing superconductivity research.

**Introduction**

Honeycomb lattices, a class of hexagonal crystal structures with corner-sharing atomic coordination, presents one of the most active fields of condensed matter physics owing to their unique electronic properties. These lattices can be constructed by either nonmetallic atoms[1-6] (e.g., in graphene and $MgB_2$) or metallic atoms[7-13] (e.g., in $Ta_4CoSi$ and $NbRh_2B_2$). Correspondingly, the electronic density of states (DOS) near the Fermi level is dominated primarily by *p*-orbitals in nonmetallic atom-based honeycomb structures, whereas *d*-orbitals dominate in their metallic counterparts.[14-16] Nevertheless, the hexagonal lattice topological symmetry of these systems gives rise to common electronic structural characteristics, including linear-dispersion near the Fermi level, highly delocalized conjugated itinerant electron states, and facilitated topological transitions through external regulation.[17-20]

Honeycomb lattices represent a promising platform for superconductivity research. Strongly correlated systems typically exhibit unconventional superconductivity, while weakly coupled systems manifest BCS-type superconductivity mediated by electron-phonon coupling. The honeycomb lattices of magic-angle graphene form a moiré superlattice that reconstructs the electronic band structure, ultimately leading to extremely narrow flat bands near the Fermi level.[21-22] Electrostatic doping overcomes the insulating gap, thereby enabling electron delocalization in the strongly correlated environment and facilitating Cooper pair formation, which induces distinct unconventional superconductivity. In the weak electron-phonon coupling regime of $MgB_2$, the boron honeycomb lattice enables $sp^2$ hybridization, which reconstructs the electronic band structure into delocalized π bands and relatively localized σ bands.[23-25] These bands enhance the DOS near the Fermi level and facilitate electron-phonon coupling. The hexagonal lattice symmetry ensures high electron mobility, which further strengthens this coupling. Such structural and electronic modulation enables Cooper pair formation, yielding BCS superconductivity with a superconducting transition temperature ($T_c$) approaching 39 K. Besides, the $CuAl_2$-type materials $Ta_4CoSi$[8, 26], $Nb_4NiSi$[9] and chiral noncentrosymmetric-type material $TaRh_2B_2$ represent another category of honeycomb network superconductor constructed by the metallic element atoms. The $T_c$ of $Ta_4CoSi$ and $Nb_4NiSi$ are 2.45 K and 7.7 K, respectively, while that of $TaRh_2B_2$ (5.8 K) falls between these two values. Interestingly, the $T_c$s of these materials are negatively correlated with the honeycomb lattices. It decreases from $MgB_2$ with the B honeycomb network to $Ta_4CoSi$ with the Ta network as the honeycomb lattice from 1.781 Å to 2.943 Å. Meanwhile, the Debye temperature decreases with the increasing of the honeycomb lattice. Besides, the shrinkage of the iron honeycomb lattice of layered $FePX_3$ (X=S and Se) under high pressure leads to the emergence of superconductivity and the increase of $T_c$.[27]

KZnBi adopts a honeycomb lattice structure formed through the alternating arrangement of Zn and Bi atoms. It crystallizes in a hexagonal ABC-type structure with the space group $P6_3/mmc$.[28] Potassium ions with large ionic radii separate adjacent planar ZnBi honeycomb layers. The alternating stacking of K ions and ZnBi honeycomb layers generates a distinct spatial symmetry. Superconductivity is observed exclusively on the (001) surface below 0.85 K under ambient pressure. Herein, we report several exotic properties of KZnBi under high pressure. Upon the application of pressure, the $T_c$ is dramatically enhanced, reaching a maximum of 7 K at 2.5 GPa. This initial increase is followed by a slight suppression of $T_c$, coinciding with a structural phase transition from the ambient-pressure $P6_3/mmc$ phase to a high-pressure $Pnma$ phase. Intriguingly, a second superconducting phase with a higher $T_c$ emerges at pressures above 7 GPa, which is correlated with an electronic transition. This work reveals a unique M-shaped double-dome superconducting phase diagram in KZnBi with honeycomb lattice under pressure, which stems from two distinct mechanisms including crystal structural phase transition and electronic transition. This finding establishes a distinctive platform for investigating the exotic superconductivity in honeycomb lattices.

**Materials and Methods**

**Sample preparation and high-pressure electrical transport measurements**

The single crystals of KZnBi utilized in this study were grown *via* a self-flux method.[28] High-pressure resistivity measurements were conducted using a nonmagnetic diamond anvil cell (DAC).[29-31] All sample storage and loading procedures were carried out in an argon-filled glovebox with oxygen and water content maintained below 0.1 ppm and 1 ppm, respectively. A cubic BN/epoxy composite layer was placed between the BeCu gasket and the electrical leads. Electrical contacts to the sample were established using four platinum (Pt) foils configured in a van der Pauw four-probe geometry for resistivity measurements. Pressure was calibrated *in situ* using the ruby luminescence method.[32] Low-temperature and magnetic field-dependent measurements were performed in a physical property measurement system (PPMS, Dynacool,

Quantum Design) with a base temperature of 1.8 K.

**High-pressure synchrotron XRD measurements**

*In situ* high-pressure X-ray diffraction (XRD) measurements were conducted at beamline 15U of the Shanghai Synchrotron Radiation Facility (SSRF), using a monochromatic X-ray beam with a wavelength of 0.6199 Å. A symmetric DAC equipped with 400 μm culet anvils and T301 stainless steel gaskets was employed in the high-pressure experiments. Silicon oil was utilized as the pressure-transmitting medium (PTM). The two-dimensional X-ray diffraction images were analyzed using the dioptas software.[33] Rietveld refinements of the crystal structures at various pressures were conducted using the General Structure Analysis System (GSAS) and graphical user interface EXPGUI package.[34]

**Theoretical calculations under varying pressures**

We conducted a comprehensive structural search on the KZnBi system at pressures of 5, 10, 15, 20 and 30 GPa using the swarm intelligence based-CALYPSO method combined with first-principle calculations.[35-37] Our first-principles calculations were carried out within the framework of density functional theory (DFT), employing the Perdew-Burke-Ernzerhof (PBE) generalized gradient approximation (GGA)[38]. These calculations were implemented using the Vienna Ab initio Simulation Package (VASP).[39] All electron projector-augmented wave (PAW) method[40] was employed, treating the $4s^1$, $3d^{10}4s^2$, and $6s^16p^3$ orbitals as valence electrons for K, Zn and Bi atoms, respectively. A plane-wave basis set with a cutoff energy of 280 eV and a Monkhorst-Pack Brillouin zone grid spacing of $2\pi \times 0.03$ Å$^{-1}$ were used to ensure convergence of enthalpies to within 1 meV/atom. Phonon dispersion calculations were performed using the frozen-phonon method as implemented in the PHONOPY package.[41] Surface states calculations were computed by constructing maximally localized Wannier functions with Wannier90[42] and subsequent analysis with WannierTools[43]. All electronic structure calculations were performed using experimentally determined lattice parameters.

## Results and Discussion

### Pressure-driven Superconductivity

We investigated the evolution of transport properties under varying pressures in KZnBi with honeycomb lattice. At ambient pressure, the resistivity $\rho(T)$ exhibits metallic behavior across the entire temperature range. Upon applying a slight pressure, a pronounced resistivity anomaly emerges in the normal state near 220 K. (Fig. 1a) This anomaly is likely linked to specific features of the electronic structure, though its precise origin remains elusive.[44-45] With further increase in pressure, the anomaly becomes progressively suppressed. Concurrently, increasing pressure leads to a continuous reduction in the overall magnitude of $\rho(T)$. Above ~3 GPa, the temperature dependence of $\rho(T)$ transitions to that characteristic of a normal metal. At ambient pressure, KZnBi undergoes a superconducting transition with a $T_c$ of approximately 0.85 K.[28] Under an applied pressure of 0.5 GPa, a slight drop in resistivity is observed at the lowest measured temperature of 1.8 K. Upon further increase in pressure to 1.2 GPa, zero resistivity is achieved at low temperatures. As illustrated in Fig. 1b, the maximum $T_c$ reaches 6.5 K at 2.6 GPa, indicating that the superconductivity in this compound is highly sensitive to hydrostatic pressure. Subsequently, the $T_c$ shifts to lower values with increasing pressure. However, an unexpected superconducting phase emerges with an onset $T_c$ of up to 7 K at 11.1 GPa. Beyond this pressure, $T_c$ decreases gradually till 32.8 GPa.

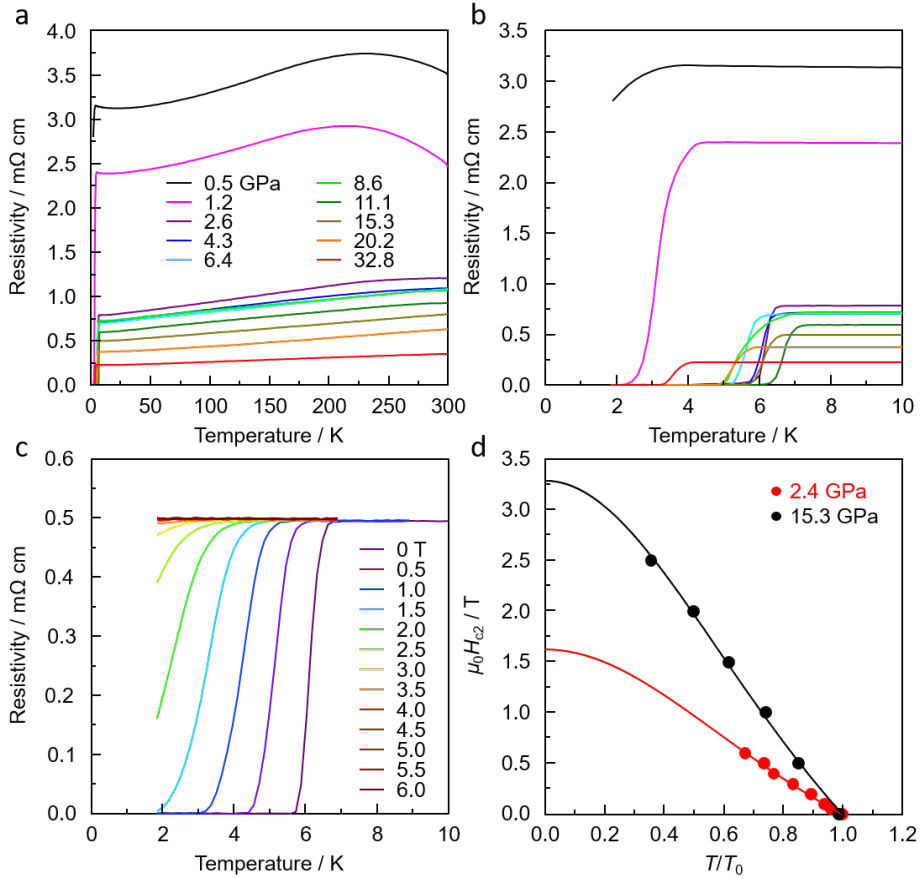

**Fig. 1. Superconductivity of KZnBi under high pressure.** (a) Temperature dependence of electrical resistivity measured at various pressures in run I; (b) Low-temperature resistivity near the superconducting $T_c$; (c) Field-dependent resistivity at 15.3 GPa, demonstrating the suppression of $T_c$ with increasing magnetic field; (d) Upper critical field for KZnBi at 15.3 GPa and 2.4 GPa. $T_c$ is defined as the temperature where resistivity drops to 10% of the normal-state value. Solid curves represent fits using the Ginzburg-Landau (GL) theory.

**Determination of Upper Critical Field**

The emergence of superconductivity is further corroborated by temperature-dependent resistivity measurements under applied magnetic fields. As illustrated in Fig. 1c, the superconducting transition shifts to lower temperatures with increasing magnetic field strength and vanishes at 4.0 T above 1.8 K under 15.3 GPa. Using the criterion of 90% of the normal-state resistivity to define the $T_c$, the temperature-magnetic field phase diagram is constructed and presented in Fig. 1d. The upper critical field, $\mu_0H_{c2}(0)$, was estimated to be 3.3 T at 15.3 GPa using the Ginzburg-Landau formula: $\mu_0H_{c2}(T) = \mu_0H_{c2}(0)(1 − t^2)/(1 + t^2)$, where t = $T/T_c$. The Ginzburg-Landau coherence length, $\xi_{GL}(0)$, was then derived from the relation $\mu_0H_{c2} = \Phi_0/(2\pi\xi^2)$, where $\Phi_0 = 2.07 \times 10^{-15}$ Wb is the magnetic flux quantum. The calculated value of $\xi_{GL}(0)$ is 10.0 nm. A similar evolution of the temperature-dependent resistivity $\rho(T)$ under applied magnetic fields is observed at 2.4 GPa. (Fig. S1) At this pressure, the $\mu_0H_{c2}(0)$ is determined to be 1.6 T, yielding a $\xi_{GL}(0)$ of 14.4 nm. Although the obtained $\mu_0H_{c2}(0)$ values remain below the corresponding Pauli paramagnetic limit, $H_P(0) = 1.84T_c$, the slopes of the upper critical field, $d\mu_0H_{c2}/dT$, exhibit a notable pressure dependence: −1.87 T/K at 2.4 GPa and −4.03 T/K at 15.3 GPa. This distinct behavior suggests that the electronic properties of the pressure-induced reentrant superconducting phase may differ fundamentally from those of the initial one.

**Transport Properties under High Pressures**

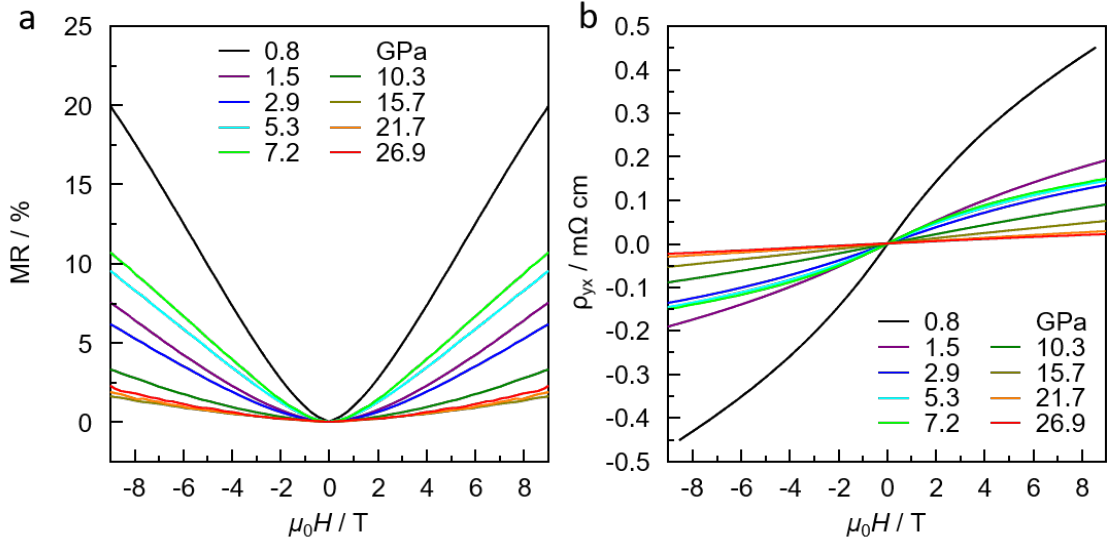

**Fig. 2. Magnetic field dependence of transport properties in KZnBi.** (a) Magnetoresistance (MR) as a function of magnetic field. (b) Hall resistivity ($\rho_{yx}$) as a function of magnetic field. The measurements were performed at 10 K under various pressures.

The superconducting properties of KZnBi demonstrate a non-monotonic evolution with increasing pressure, suggesting the emergence of an unconventional electronic structure under compression. Further evidence of anomalous behavior is observed in field-dependent resistivity measurements conducted at high pressures. Fig. 2 depicts the field dependence of magnetoresistance (MR) and Hall resistivity $\rho_{yx}$ at 10 K under various pressures. Although the MR remains positive throughout the measured pressure range (Fig. 2a), its magnitude shows a non-monotonic evolution, characterized by an initial decrease, followed by an increase, and a subsequent decrease with increasing pressure. Moreover, Hall effect measurements reveal that $\rho_{yx}$ exhibits a non-linear dependence on the magnetic field (B), as shown in Fig. 2b. This non-linearity is more pronounced at lower pressures. The deviation from linearity indicates that the charge transport cannot be described by a simple single-band model. Consequently, we employed a two-carrier model to analyze the data.

$$\rho_{yx} = \frac{B}{e} \frac{(\mu_h^2 n_h - \mu_e^2 n_e) + (\mu_h \mu_e)^2 B^2 (n_h - n_e)}{(\mu_h n_h + \mu_e n_e)^2 + (\mu_h \mu_e)^2 B^2 (n_h - n_e)^2} \tag{1}$$

All $\rho_{yx}(B)$ data collected below 7 GPa were successfully fitted using a two-band model, where the carrier concentrations ($n_e$, $n_h$) and mobilities ($\mu_e$, $\mu_h$) were treated as free parameters. The pressure dependencies of the electron and hole concentrations extracted from these fits are displayed in Fig. 5. With pressure increasing, both the electron and hole carrier concentrations increase and the peak at $P_c$ of 2.9 GPa coincides with the maximum $T_c$. At pressures exceeding 7 GPa, the Hall resistivity data signify an electronic transition, marked by a shift from multi-band charge transport to purely hole-type conduction. This experimental observation corroborates the evolution of the Fermi surface, which aligns with the predictions from our theoretical calculations.

**Structure Evolution under High Pressures**

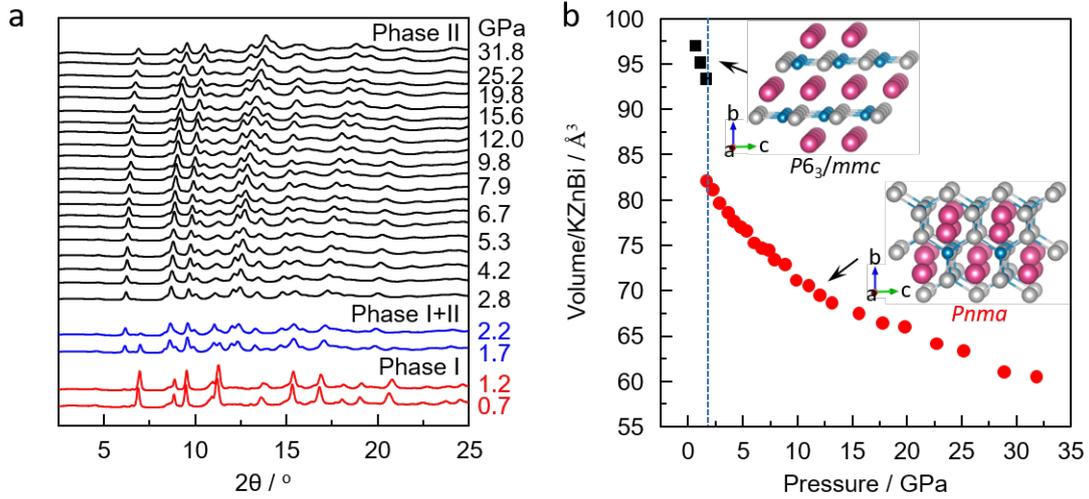

**Fig. 3. Structural characterization of KZnBi under high pressure.** (a) X-ray diffraction patterns obtained at room temperature under pressures up to 31.8 GPa (λ = 0.6199 Å). The color-coded patterns (red, blue, black) indicate distinct phase transitions during compression; (b) Pressure dependence of unit-cell volume, with solid lines representing Birch-Murnaghan equation of state fits. The inset illustrates the crystal structures of ambient and high-pressure phase.

To understand the effect of pressure on the structure and to uncover the key factor controlling superconductivity, we performed *in situ* high-pressure synchrotron XRD on KZnBi. As shown in Fig. 3, the sample retains its ambient-pressure hexagonal structure (space group *P*6$_3$/*mmc*, No. 194) across the low-pressure range. A representative Rietveld refinement at 0.7 GPa is presented in Fig. S2a. All observed diffraction peaks are successfully indexed by this structural model. However, a set of new diffraction peaks emerges at 1.7 GPa and becomes dominant upon further compression, clearly indicating a structural phase transition.

To identify the high-pressure phase of KZnBi, we performed extensive structure searches using the state-of-the-art CALYPSO method combined with first-principles DFT calculations[36-37]. A new ground state with the space group *Pnma* emerges at 1.5 GPa. This phase transition pressure is in good agreement with experimental observations. The predicted crystal structure for the high-pressure phase of KZnBi is shown inset of Fig. 3b. The ambient-pressure phase features a quasi-two-dimensional structure, consisting of Zn-Bi honeycomb layers intercalated by K atoms. In contrast, the high-pressure phase adopts a 3D network formed by Zn and Bi atoms, with K atoms occupying the interstitial sites. As pressure increases, the high-pressure phase remains on the ternary convex hull and becomes energetically more favorable relative to the ambient-pressure phase (Fig. S3). We also calculated the phonon spectrum. The absence of imaginary frequencies confirms its dynamic stability under pressure (Fig. S4). The excellent Rietveld refinements achieved using the predicted structures provide definitive confirmation of the high-pressure phase's structure (Fig. S2b). The pressure dependence of the lattice parameters (Fig. S5) and the unit cell volume (Fig. 3b) exhibits a distinct discontinuity around 2.8 GPa in KZnBi.

**First-principles Calculations of Electronic Band Structure**

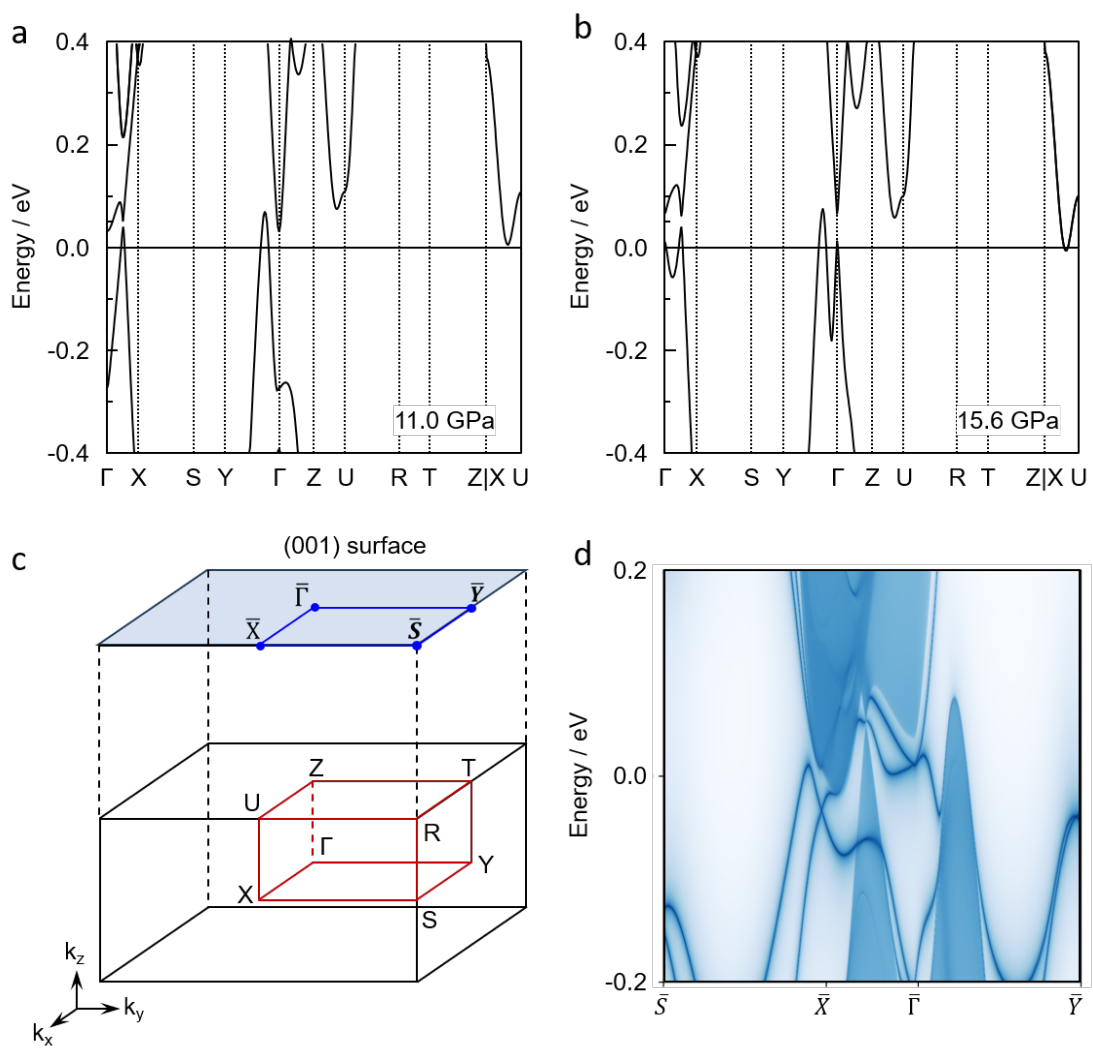

**Fig. 4. Electronic structure of KZnBi under high pressure.** (a) Calculated band structure of the *Pnma* phase at 11.0 GPa. (b) Band structure evolution at 15.6 GPa, showing pressure-induced modifications. (c) Bulk Brillouin zone of the *Pnma* phase

and its corresponding (001) surface projection. (d) Topological surface states calculated for the (001) surface at 11.0 GPa.

Previous studies have reported that KZnBi at ambient pressure exhibits bulk Dirac cone states and topological surface states on the (010) surface. To gain further insight into the topological properties of its high-pressure phase, we calculated its electronic band structure under pressure. As shown in Fig. 4, the high-pressure phase of KZnBi exhibits semi-metallic behavior characterized by a continuous band gap, though it lacks the 3D Dirac cones observed at ambient pressure. The electronic band structure remains largely unchanged below 15.6 GPa. At this critical pressure, however, an emergent hole pocket appears near the Gamma point, resulting in a pressure-induced electronic transition. This feature suggests a potential pressure-induced alteration of the topological properties. The calculated $\mathbb{Z}_2$ invariant for high-pressure phase is $(\upsilon_0 : \upsilon_1\upsilon_2\upsilon_3)$ = (1 : 111), confirming its classification as a strong topological semimetal (see Fig. S6 and S7). Fig. 4d displays the surface states on the (001) surface at 11.0 GPa, revealing a Dirac cone located at the $\bar{X}$ point just below the Fermi level. As anticipated, at 15.6 GPa, the topological surface states undergo a transformation: a new Dirac point emerges at the $\Gamma$ point above the Fermi level, while the Dirac cone at the X point shifts further below it (Fig. S8). Therefore, our combined theoretical calculations and *in situ* high-pressure measurements demonstrate the coexistence of topological electronic states and superconductivity in KZnBi under compression.

**Phase Diagram under High Pressure**

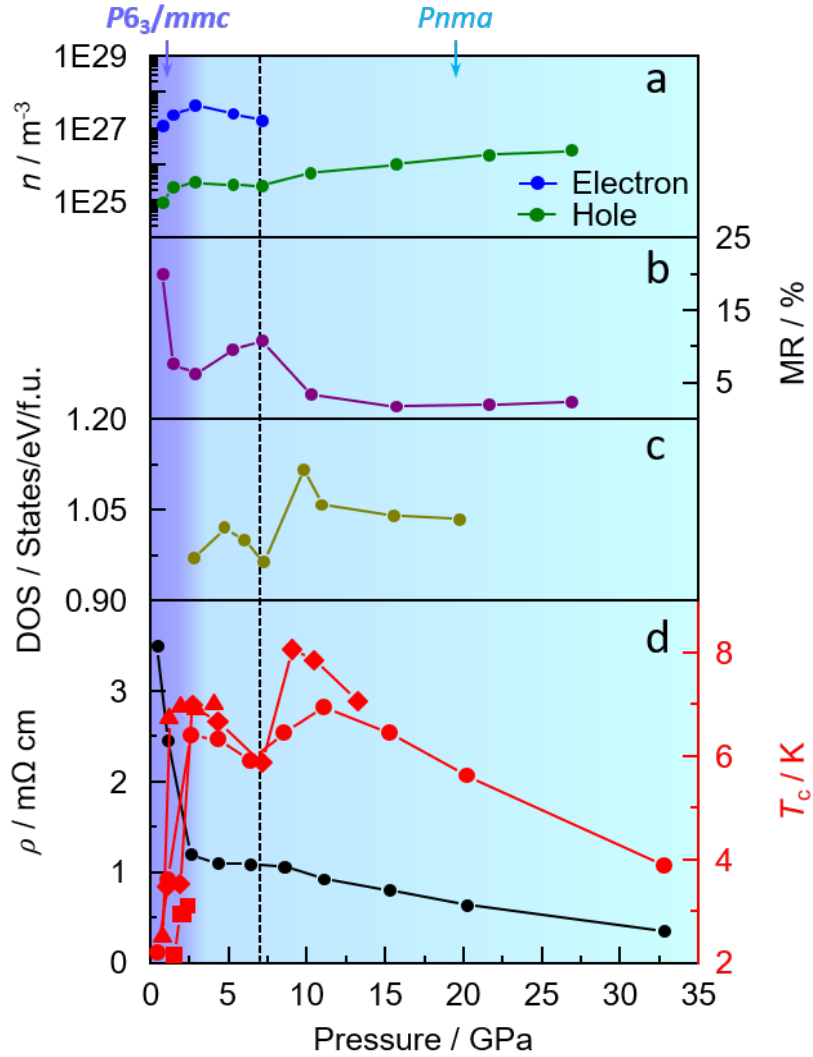

**Fig. 5. Phase diagram of KZnBi under high pressures.** (a) Evolution of carrier concentrations showing hole (olive) and electron (blue) contributions determined by Hall effect measurements. (b) Magnetoresistance behavior under applied pressures. (c) Pressure dependence of the density of states (DOS, yellow). (d) Pressure variation of superconducting transition temperatures ($T_c$, red, circles, triangles, diamonds and squares represent Run I, Run II, Run III and Run IV, respectively), with onset values extracted from resistivity measurements. The values of room-temperature resistivity (black) were also shown here.

Based on the high-pressure experimental results presented above, we constructed the temperature-pressure phase diagram of KZnBi, as shown in Fig. 5. High-pressure transport measurements performed on KZnBi single crystals across multiple independent experimental runs yielded consistent and reproducible results, confirming the intrinsic nature of the observed superconductivity under pressure (Fig. S9-S11). The application of high pressure dramatically alters both the electronic and crystal structures of KZnBi. Both electrical resistivity and magnetoresistance exhibit a sharp decrease with increasing pressure until a structural phase transition from the *P*6$_3$/*mmc* to the *Pnma* space group occurs at ~2 GPa. Beyond this transition pressure, the resistivity declines more gradually, while the magnetoresistance exhibits a slow but distinct increase. Concurrently, the $T_c$ increases significantly, reaching a maximum of 7 K at approximately 3 GPa. Subsequently, $T_c$ begins to decline, a suppression that coincides with the structural phase transition. Within this pressure range, charge transport involves contributions from both electrons and holes, with electrons initially serving as the majority carriers. The increase of carrier concentration may thus contribute to the initial increase of $T_c$. However, once the pressure surpasses 7.2 GPa, the linear magnetic field dependence of the Hall resistivity ($\rho_{yx}$) indicates a shift in carrier dominance toward holes, concurrent with the onset of a secondary reduction in magnetoresistance. Remarkably, the re-emergence of superconductivity coincides with this transition in carrier dominance, resulting in a new superconducting phase that exhibits an elevated $T_c$. Our results indicate that the re-entrant superconductivity can be attributed to a pressure-induced electronic transition, which is further supported by our theoretical calculations. Subsequently, a double dome-shaped superconducting phase emerges. The experimentally observed variation in the $T_c$ can be explained by the calculated DOS at the Fermi level, which also exhibits an approximately M-shaped evolution under pressure (Fig. 5). This correlation suggests that the $T_c$ of KZnBi is highly sensitive to the DOS value at the Fermi level.

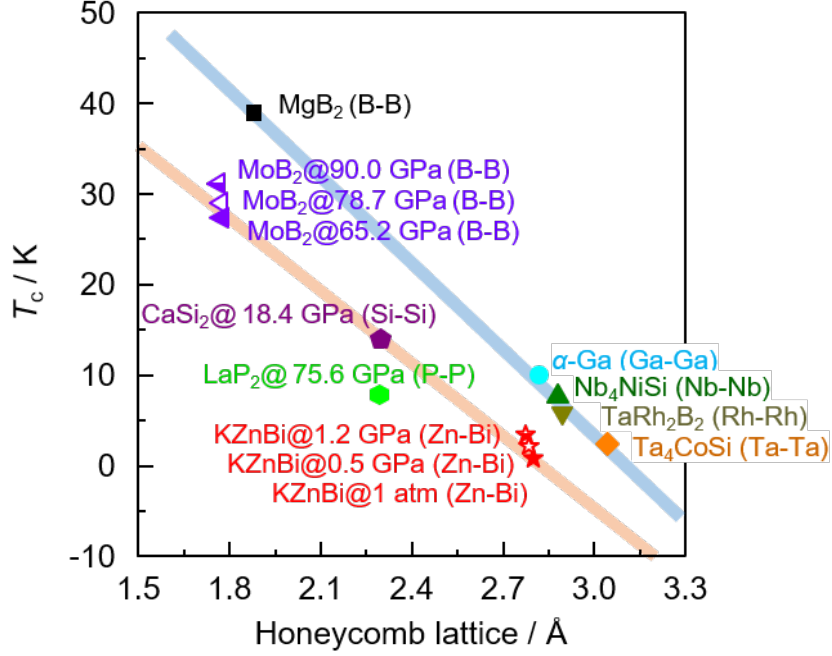

**Fig. 6. The honeycomb lattice parameter dependence of $T_c$.**[3, 8-9, 13, 25, 30, 46-47] The blue and orange shaded bands act as visual guides to highlight the distinct trends.

Finally, we address the critical relationship between the $T_c$ and honeycomb lattice geometry. Fig. 6 reveals an approximately inverse linear correlation between $T_c$ and the honeycomb lattice parameter for KZnBi and representative honeycomb superconductors. At ambient pressure, $T_c$ systematically increases as the honeycomb lattice decreases, spanning from $Ta_4CoSi$[8] (metallic Ta-Ta network ≈ 2.943 Å) to $MgB_2$[3] (covalent B-B network ≈ 1.781 Å). Remarkably, this trend persists under high pressure, where the honeycomb lattice compression universally enhances $T_c$ across multiple honeycomb systems. (Table S1) The robustness of this correlation underscores the geometric control of superconductivity within the honeycomb layer and establishes lattice compression as a key design principle for discovering higher $T_c$ superconductors with a honeycomb network.

**Conclusion**

In summary, high pressure serves as a clean tuning parameter for systematically investigating the evolution of structural and electrical properties in KZnBi with honeycomb lattice. Our study reveals that applied pressure significantly enhances the $T_c$ until a structural phase transition occurs around 2 GPa, beyond which $T_c$ gradually decreases. Further compression induces an electronic transition that modifies the electronic structure and leads to the re-emergence of superconductivity with a higher $T_c$. Interestingly, theoretical calculations indicate that the high-pressure phase of KZnBi is a topological metal, coexisting with superconductivity. Consequently, our work establishes KZnBi as an ideal prototype system for investigating emergent quantum phenomena in honeycomb lattices and exploring pressure-controlled superconductivity.

## Acknowledgements

This work was supported by the National Natural Science Foundation of China (Grant No. 52272265 and 12474018), the National Key R&D Program of China (Grant No. 2023YFA1607400) and the National Research Foundation of Korea (NRF) grant funded by the Korea government (MSIT) (Grant No. 2022R1A2C2004735). The authors thank the Analytical Instrumentation Center (# SPST-AIC10112914), SPST, ShanghaiTech University for assistance in facility support. The authors thank the staffs from BL15U1 at Shanghai Synchrotron Radiation Facility for assistance during data collection.

## Conflict of Interest

The authors declare no competing interests.

## Supporting Information

The Supporting Information contains transport results of KZnBi in run II and run III, XRD pattern in decompression, pressure dependence of lattice parameter and typical Rietveld refinement of KZnBi in two phases.

## References

1. Novoselov, K. S.; Geim, A. K.; Morozov, S. V.; Jiang, D.; Zhang, Y.; Dubonos, S. V.; Grigorieva, I. V.&Firsov, A. A., Electric field effect in atomically thin carbon films. *Science* 306, 666 (2004).
2. Geim, A. K.&Novoselov, K. S., The rise of graphene. *Nat. Mater.* 6, 183-191 (2007).
3. Nagamatsu, J.; Nakagawa, N.; Muranaka, T.; Zenitani, Y.&Akimitsu, J., Superconductivity at 39 K in magnesium diboride. *Nature* 410, 63-64 (2001).
4. Choi, H. J.; Roundy, D.; Sun, H.; Cohen, M. L.&Louie, S. G., The origin of the anomalous superconducting properties of MgB$_2$. *Nature* 418, 758-760 (2002).

5. Tao, L.; Cinquanta, E.; Chiappe, D.; Grazianetti, C.; Fanciulli, M.; Dubey, M.; Molle, A.&Akinwande, D., Silicene field-effect transistors operating at room temperature. *Nat. Nanotechnol.* 10, 227-231 (2015).
6. Po, H. C.; Zou, L.; Vishwanath, A.&Senthil, T., Origin of Mott insulating behavior and superconductivity in twisted bilayer graphene. *Phys. Rev. X* 8, 031089 (2018).
7. Singh, Y.&Gegenwart, P., Antiferromagnetic Mott insulating state in single crystals of the honeycomb lattice material $Na_2IrO_3$. *Phys. Rev. B* 82, 064412 (2010).
8. Zeng, L.; Hu, X.; Guo, S.; Lin, G.; Song, J.; Li, K.; He, Y.; Huang, Y.; Zhang, C.; Yu, P.; Ma, J.; Yao, D.-X.&Luo, H., $Ta_4CoSi$: A tantalum-rich superconductor with a honeycomb network structure. *Phys. Rev. B* 106, 134501 (2022).
9. Ryu, G.; Kim, S. W.; Matsuishi, S.; Kawaji, H.&Hosono, H., Superconductivity in $Nb_4MSi$ (M=Ni, Co, and Fe) with a quasi-two-dimensional Nb network. *Phys. Rev. B* 84, 224518 (2011).
10. Viciu, L.; Huang, Q.; Morosan, E.; Zandbergen, H. W.; Greenbaum, N. I.; McQueen, T.&Cava, R. J., Structure and basic magnetic properties of the honeycomb lattice compounds $Na_2Co_2TeO_6$ and $Na_3Co_2SbO_6$. *J. Solid State Chem.* 180, 1060-1067 (2007).
11. Nishikubo, Y.; Kudo, K.&Nohara, M., Superconductivity in the honeycomb-lattice pnictide SrPtAs. *J. Phys. Soc. Jpn.* 80, 055002 (2011).
12. Long, G.; Zhang, T.; Cai, X.; Hu, J.; Cho, C. W.; Xu, S.; Shen, J.; Wu, Z.; Han, T.; Lin, J.; Wang, J.; Cai, Y.; Lortz, R.; Mao, Z.&Wang, N., Isolation and characterization of few-layer manganese thiophosphite. *ACS Nano* 11, 11330-11336 (2017).
13. Carnicom, E. M.; Xie, W.; Klimczuk, T.; Lin, J.&Cava, R. J., $TaRh_2B_2$ and $NbRh_2B_2$: Superconductors with a chiral noncentrosymmetric crystal structure. *Sci. Adv.* 4, eaar7969 (2018).
14. An, J. M.&Pickett, W. E., Superconductivity of $MgB_2$: covalent bonds driven metallic. *Phys. Rev. Lett.* 86, 4366-4369 (2001).
15. Zeng, H.; Dai, J.; Yao, W.; Xiao, D.&Cui, X., Valley polarization in $MoS_2$ monolayers by optical pumping. *Nat. Nanotechnol.* 7, 490-493 (2012).
16. Li, L.; Wang, Y.; Xie, S.; Li, X. B.; Wang, Y. Q.; Wu, R.; Sun, H.; Zhang, S.&Gao, H. J., Two-dimensional transition metal honeycomb realized: Hf on Ir(111). *Nano Lett* 13, 4671-4674 (2013).
17. Wu, W.; Chen, Y.-H.; Tao, H.-S.; Tong, N.-H.&Liu, W.-M., Interacting Dirac fermions on honeycomb lattice. *Phys. Rev. B* 82, 245102 (2010).
18. Wu, L. H.&Hu, X., Topological properties of electrons in honeycomb lattice with detuned hopping energy. *Sci Rep* 6, 24347 (2016).
19. Das Sarma, S.; Adam, S.; Hwang, E. H.&Rossi, E., Electronic transport in two-dimensional graphene. *Rev. Mod. Phys.* 83, 407-470 (2011).
20. Peres, N. M. R., Colloquium: The transport properties of graphene: An introduction. *Rev. Mod. Phys.* 82, 2673-2700 (2010).
21. Cao, Y.; Fatemi, V.; Fang, S.; Watanabe, K.; Taniguchi, T.; Kaxiras, E.&Jarillo-Herrero, P., Unconventional superconductivity in magic-angle graphene superlattices. *Nature* 556, 43-50 (2018).
22. Uchoa, B.&Castro Neto, A. H., Superconducting states of pure and doped graphene. *Phys. Rev. Lett.* 98, 146801 (2007).
23. Kortus, J.; Mazin, I. I.; Belashchenko, K. D.; Antropov, V. P.&Boyer, L. L., Superconductivity of metallic boron in $MgB_2$. *Phys. Rev. Lett.* 86, 4656-4659 (2001).
24. Buzea, C.&Yamashita, T., Review of the superconducting properties of $MgB_2$. *Supercond. Sci.*

Technol. 14, R115-R146 (2001).

25. Zhang, M.; Pei, C.; Zhu, B.; Wang, Q.; Wu, J.&Qi, Y., Pressure-induced superconductivity in LaP$_2$ with a graphenelike phosphorus layer. *Phys. Rev. B* 112, 184108 (2025).

26. Bekaert, J., Phonon-mediated superconductivity in ternary silicides X$_4$CoSi (X=Nb, Ta). *Phys. Rev. B* 108, 134504 (2023).

27. Wang, Y.; Ying, J.; Zhou, Z.; Sun, J.; Wen, T.; Zhou, Y.; Li, N.; Zhang, Q.; Han, F.; Xiao, Y.; Chow, P.; Yang, W.; Struzhkin, V. V.; Zhao, Y.&Mao, H. K., Emergent superconductivity in an iron-based honeycomb lattice initiated by pressure-driven spin-crossover. *Nat. Commun.* 9, 1914 (2018).

28. Song, J.; Kim, S.; Kim, Y.; Fu, H.; Koo, J.; Wang, Z.; Lee, G.; Lee, J.; Oh, S. H.; Bang, J.; Matsushita, T.; Wada, N.; Ikegami, H.; Denlinger, J. D.; Lee, Y. H.; Yan, B.; Kim, Y.&Kim, S. W., Coexistence of surface superconducting and three-dimensional topological Dirac states in semimetal KZnBi. *Phys. Rev. X* 11, 021065 (2021).

29. Pei, C.; Ying, T.; Zhang, Q.; Wu, X.; Yu, T.; Zhao, Y.; Gao, L.; Li, C.; Cao, W.; Zhang, Q.; Schnyder, A. P.; Gu, L.; Chen, X.; Hosono, H.&Qi, Y., Caging-Pnictogen-Induced superconductivity in skutterudites IrX$_3$ (X = As, P). *J. Am. Chem. Soc.* 144, 6208-6214 (2022).

30. Pei, C.; Zhang, J.; Wang, Q.; Zhao, Y.; Gao, L.; Gong, C.; Tian, S.; Luo, R.; Li, M.; Yang, W.; Lu, Z.-Y.; Lei, H.; Liu, K.&Qi, Y., Pressure induced superconductivity at 32 K in MoB$_2$. *Natl. Sci. Rev.* 10, nwad034 (2023).

31. Pei, C.; Zhang, M.; Peng, D.; Huangfu, S.; Zhu, S.; Wang, Q.; Wu, J.; Xing, Z.; Zhang, L.; Chen, Y.; Zhao, J.; Yang, W.; Suo, H.; Guo, H.; Zeng, Q.&Qi, Y., Unveiling pressurized bulk superconductivity in a trilayer nickelate Pr$_4$Ni$_3$O$_{10}$ single crystal. *Sci. China-Phys. Mech. Astron.*, (2025).

32. Mao, H. K.; Xu, J.&Bell, P. M., Calibration of the ruby pressure gauge to 800 kbar under quasi-hydrostatic conditions. *J. Geophys. Res.* 91, 4673-4676 (1986).

33. Prescher, C.&Prakapenka, V. B., DIOPTAS: a program for reduction of two-dimensional X-ray diffraction data and data exploration. *High Pressure Res.* 35, 223-230 (2015).

34. Larson, A. C.&Dreele, R. B. V., General structure analysis system (GSAS). *Los Alamos National Laboratory Report LAUR* 86-748 (2004).

35. Gao, B.; Gao, P.; Lu, S.; Lv, J.; Wang, Y.&Ma, Y., Interface structure prediction via CALYPSO method. *Sci. Bull.* 64, 301-309 (2019).

36. Wang, Y.; Lv, J.; Zhu, L.&Ma, Y., Crystal structure prediction via particle-swarm optimization. *Phys. Rev. B* 82, 094116 (2010).

37. Wang, Y.; Lv, J.; Zhu, L.&Ma, Y., CALYPSO: A method for crystal structure prediction. *Comput. Phys. Commun.* 183, 2063-2070 (2012).

38. Perdew, J. P.; Burke, K.&Ernzerhof, M., Generalized gradient approximation made simple. *Phys. Rev. Lett.* 77, 3865-3868 (1996).

39. Kresse, G.&Furthmüller, J., Effcient iterative schemes for ab initio total-energy calculations. *Phys. Rev. B* 54, 11169 (1996).

40. Blochl, P. E., Projector augmented-wave method. *Phys. Rev. B* 50, 17953-17979 (1994).

41. Togo, A.; Oba, F.&Tanaka, I., First-principles calculations of the ferroelastic transition between rutile-type andCaCl$_2$-type SiO$_2$ at high pressures. *Phys. Rev. B* 78, 134106 (2008).

42. Pizzi, G.; Vitale, V.; Arita, R.; Blügel, S.; Freimuth, F.; Géranton, G.; Gibertini, M.; Gresch, D.; Johnson, C.; Koretsune, T.; Ibañez-Azpiroz, J.; Lee, H.; Lihm, J.-M.; Marchand, D.; Marrazzo, A.; Mokrousov, Y.; Mustafa, J. I.; Nohara, Y.; Nomura, Y.; Paulatto, L.; Poncé, S.; Ponweiser, T.; Qiao, J.;

Thöle, F.; Tsirkin, S. S.; Wierzbowska, M.; Marzari, N.; Vanderbilt, D.; Souza, I.; Mostofi, A. A.&Yates, J. R., Wannier90 as a community code: new features and applications. *J. Phys.: Condens. Matter* 32, 165902 (2020).

43. Wu, Q.; Zhang, S.; Song, H.-F.; Troyer, M.&Soluyanov, A. A., WannierTools: An open-source software package for novel topological materials. *Comput. Phys. Commun.* 224, 405-416 (2018).

44. Qi, Y.; Shi, W.; Naumov, P. G.; Kumar, N.; Schnelle, W.; Barkalov, O.; Shekhar, C.; Borrmann, H.; Felser, C.; Yan, B.&Medvedev, S. A., Pressure dirven superconductivity in the transition metal pentatelluride $HfTe_5$. *Phys. Rev. B* 94, 054517 (2016).

45. McIlroy, D. N.; Moore, S.; Zhang, D.; Wharton, J.; Kempton, B.; Littleton, R.; Wilson, M.; Tritt, T. M.&Olson, C. G., Observation of a semimetal–semiconductor phase transition in the intermetallic $ZrTe_5$. *J. Phys.: Condens. Matter* 16, L359-L365 (2004).

46. Bordet, P.; Affronte, M.; Sanfilippo, S.; Núñez-Regueiro, M.; Laborde, O.; Olcese, G. L.; Palenzona, A.; LeFloch, S.; Levy, D.&Hanfland, M., Structural phase transitions in $CaSi_2$ under high pressure. *Phys. Rev. B* 62, 11392 (2000).

47. Petrov, M.; Bekaert, J.&Milošević, M. V., Superconductivity in gallenene. *2D Materials* 8, 035056 (2021).